\documentclass[english,pre, superscriptaddress, longbibliography, preprint]{revtex4-1}
\usepackage[T1]{fontenc}
\usepackage[latin9]{inputenc}
\usepackage{color}
\usepackage{bm}
\usepackage{amsmath}
\usepackage{amssymb}
\usepackage{graphicx}

\makeatletter
\usepackage{amsmath}
\usepackage{subfigure}
\usepackage{graphicx, mathtools}
\usepackage[normalem]{ulem}
\usepackage[colorlinks=true,linkcolor=MidnightBlue,urlcolor=MidnightBlue,citecolor=MidnightBlue,anchorcolor=MidnightBlue]{hyperref}\usepackage[dvipsnames]{xcolor}
\usepackage{chngcntr}

\makeatother

\usepackage{babel}
\begin{document}

\title{Competing chemical and hydrodynamic interactions in autophoretic
colloidal suspensions}
\begin{abstract}
At the surfaces of autophoretic colloids, slip velocities arise from
local chemical gradients that are many-body functions of particle
configuration and activity. For rapid chemical diffusion, coupled
with slip-induced hydrodynamic interactions, we deduce the chemohydrodynamic
forces and torques between colloids. For bottom-heavy particles above
a no-slip wall, the forces can be expressed as gradients of a non-equilibrium
potential which, by tuning the type of activity, can be varied from
repulsive to attractive. When this potential has a barrier, we find
arrested phase separation with a mean cluster size set by competing
chemical and hydrodynamic interactions. These are controlled, in turn,
by the monopolar and dipolar contributions to the active chemical
surface fluxes. 
\end{abstract}

\author{Rajesh Singh}
\email{rs2004@cam.ac.uk}

\selectlanguage{english}%

\affiliation{DAMTP, Centre for Mathematical Sciences, University of Cambridge,
Wilberforce Road, Cambridge CB3 0WA, UK}

\author{R. Adhikari}

\affiliation{DAMTP, Centre for Mathematical Sciences, University of Cambridge,
Wilberforce Road, Cambridge CB3 0WA, UK}

\affiliation{The Institute of Mathematical Sciences-HBNI, CIT Campus, Chennai
600113, India}

\author{M. E. Cates}

\affiliation{DAMTP, Centre for Mathematical Sciences, University of Cambridge,
Wilberforce Road, Cambridge CB3 0WA, UK}
\maketitle

\section{Introduction}

Non-equilibrium processes, of biological \cite{brennen1977} or chemical
\cite{anderson1989colloid,ebbens2010pursuit} origin, when confined
to a thin layer around a colloidal particle, create interfacial slip
flows. These drive exterior fluid flows, which mediate long-ranged
hydrodynamic interactions between particles \cite{singh2015many,singh2016crystallization,singh2018generalized,thutupalli2018FIPS}.
In instances where the slip is of biological origin, as for example
in suspensions of \textit{Volvox} \cite{goldstein2015green,pedley2016spherical},
the slip on any one particle is typically independent of the configuration
and slip of other particles. Motion under the active hydrodynamic
forces and torques thereby computed \cite{singh2015many,singh2016crystallization,singh2018generalized}
is in excellent agreement with experiment \cite{petroff2015fast,thutupalli2018FIPS}.
However, when slip is induced by gradients of chemical species, as
for example in autophoretic colloids \cite{palacci2013living,ebbens2010pursuit},
the gradient at the location of one particle is determined by the
chemical fields of all other particles, causing the slip to be a many-body
function of the colloidal configurations and activities. Interactions
in autophoretic suspensions thus have both chemical and hydrodynamic
many-body contributions. A quantitative theory of these is necessary
to understand the dynamics and non-equilibrium steady-states of such
suspensions.

Experiments on active suspensions are often performed in the vicinity
of a plane boundary, for example a no-slip wall \cite{theurkauff2012dynamic,buttinoni2013DynamicClustering,palacci2013living}.
Here, aggregation of colloids to a self-limiting cluster size that
is proportional to the self-propulsion speed has been reported. This
cannot be explained by current theories \cite{pohl2014dynamicClustering,sahaPRE2014,liebchen2017phoretic,yan2016behavior},
which typically only account for chemical (not hydrodynamic) many-body
effects. (Many-body chemohydrodynamics have recently been addressed,
but not near a wall \cite{liebchen2018interactions,popescu2018effective}.)
While it has been shown that translational and rotational diffusiophoretic
motion in overlapping chemical fields can induce aggregation \cite{pohl2014dynamicClustering},
this predicts instead a decrease in the size of aggregates with self-propulsion
speed \cite{theurkauff2012dynamic,buttinoni2013DynamicClustering}.
In other work, aggregation has been attributed to the tendency of
particles to propel away from the chemical they produce \cite{liebchen2015clustering,liebchen2017phoretic}
leading to formation of patterns. So far, all of the above theories
ignore the local conservation of momentum and/or the role played by
plane boundaries as barriers to chemical flux and sinks of fluid momentum.
A theory which consistently accounts for these effects remains lacking. 

Here we present a microscopic theory for autophoretic colloids in
the proximity of boundaries, restricting attention, for simplicity,
to cases governed by a single chemical diffusant species. We construct
the slip on one particle as a many-body function of the position and
activity of all others. The slip is obtained from the solution of
the diffusion equation, in the limit of zero Péclet number, and then
used in the momentum equation, in the limit of zero Reynolds number,
to compute chemohydrodynamic forces and torques between the colloids. 

These forces and torques generically do not admit potentials and explicitly
violate the action-reaction principle \cite{soto2014self}. However,
for bottom-heavy particles located above and oriented normal to a
plane boundary where the chemical flux and fluid flow vanish, the
bulk flow is predominantly irrotational and, strikingly, the interactions
can after all be written as gradients of a non-equilibrium pair potential.
To leading order, this potential depends on the ratio, $\alpha$,
of the magnitudes of monopolar and dipolar chemical activity of the
colloids (see below) and can be varied, by tuning this ratio, from
purely repulsive to purely attractive. This leads to colloidal steady-states
that are, respectively, liquid-like and crystalline. When instead
the potential has a barrier, phase separation can arrest to a self-limiting
cluster size that is set by $\alpha$.

In what follows, we explain how these results are derived, starting
in Section \ref{MBC} with our solution of the chemohydrodynamic traction
in the presence of fast diffusant. Section \ref{sec:minimalModel}
applies our formalism to the case of bottom-heavy particles near an
infinite plane horizontal no-flux, no-slip wall and we conclude with
a brief discussion in Section \ref{sec:disc}.

\section{Many-body Chemohydrodynamics\label{MBC}}

\begin{figure*}[t]
\centering\includegraphics[width=0.9\textwidth]{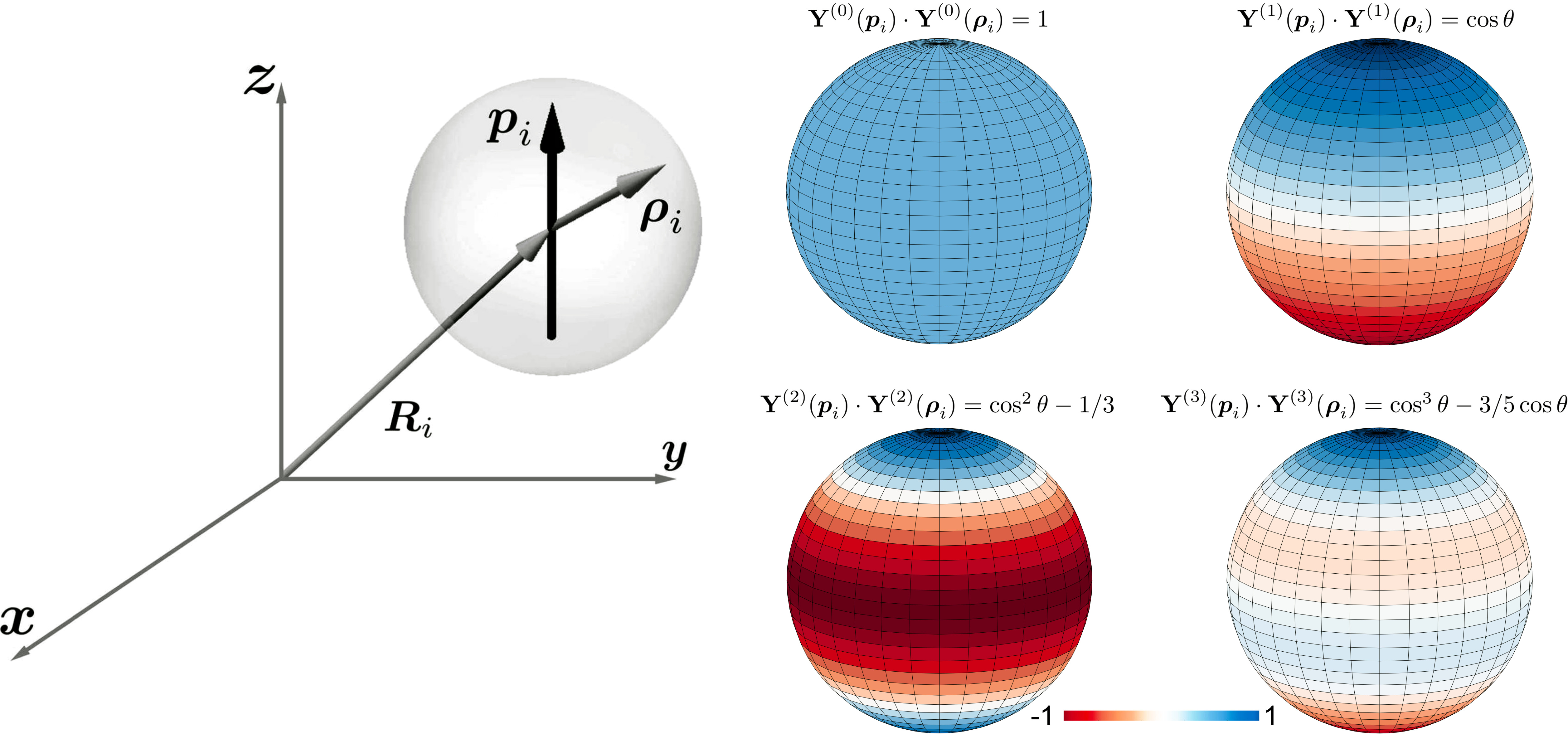}\caption{The left panel shows the system of coordinates used to describe the
kinematics of active colloids. The $i$-th colloid is centered at
$\boldsymbol{R}_{i}$, oriented along the unit vector $\boldsymbol{p}_{i}$,
and has radius vector $\boldsymbol{\rho}_{i}$. The right panel shows
the tensorial modes, $\mathbf{Y}_{i}^{(m)}(\bm{p}_{i})\cdot\mathbf{Y}^{(m)}(\bm{\hat{\rho}}_{i})$,
of an uniaxial surface scalar field with the direction of axial symmetry
along the $z$-axis, for $m=0,\ldots3$.\label{fig:1chem}}
\end{figure*}
We consider a suspension of $N$ chemically active spherical colloids
of radius $b$ in an incompressible fluid of viscosity $\eta$. The
$i$-th sphere is centered at $\boldsymbol{R}_{i}$, has radius vector
$\boldsymbol{\rho}_{i}$, and is oriented along $\boldsymbol{p}_{i}$
with the points $\boldsymbol{r}_{i}=\boldsymbol{R}_{i}+\boldsymbol{\rho}_{i}$
defining its surface $S_{i}$. The system of coordinates is shown
in Fig.(\ref{fig:1chem}). Here we restrict ourselves to the limit
of rapid Fickian diffusion of chemical and rapid viscous transport
of momentum. Then, the chemical field $c$ obeys the steady-state
diffusion equation $\boldsymbol{\nabla}\cdot\boldsymbol{j}=0$ in
the bulk where $\boldsymbol{j}=-D\boldsymbol{\nabla}c$ is the diffusive
flux with diffusivity $D$. The chemical reaction determines the normal
component of the flux at the boundaries,
\begin{equation}
\boldsymbol{j}(\boldsymbol{r})\cdot\hat{\boldsymbol{\rho}_{i}}=j^{\mathcal{A}}(\boldsymbol{\rho}_{i})\qquad(\boldsymbol{r}\in S_{i}).
\end{equation}
Here and below, fields that are restricted to the surfaces $S_{i}$
carry arguments $\boldsymbol{\rho}_{i}$. The slip flow produced by
chemical gradients is
\begin{alignat}{1}
\boldsymbol{v}^{\mathcal{A}}(\boldsymbol{\rho}_{i}) & =\mu_{c}(\boldsymbol{\rho}_{i})\boldsymbol{\nabla}_{s}\,c(\boldsymbol{\rho}_{i}),\label{eq:slipConcGrad}
\end{alignat}
where $\boldsymbol{\nabla}_{s}$ is the surface gradient and $\mu_{c}(\boldsymbol{\rho}_{i})$
is the phoretic mobility \cite{anderson1989colloid}. The flow field
$\boldsymbol{v}$ obeys the momentum balance equation $\boldsymbol{\nabla}\cdot\boldsymbol{\sigma}=0$
where $\boldsymbol{\sigma}=-p\boldsymbol{I}+\eta(\bm{\nabla}\boldsymbol{v}+(\bm{\nabla}\boldsymbol{v})^{T})$
is the Cauchy stress in an incompressible Newtonian fluid and $\boldsymbol{I}$
is the identity tensor. The slip flow modifies the usual no-slip boundary
condition on a colloid translating with velocity $\mathbf{V}_{i}$
and rotating with angular velocity $\mathbf{\Omega}_{i}$ to
\begin{alignat}{1}
\boldsymbol{v}(\boldsymbol{r}) & =\mathbf{V}_{i}+\mathbf{\Omega}_{i}\times\boldsymbol{\rho}_{i}+\boldsymbol{v}^{\mathcal{A}}(\boldsymbol{\rho}_{i})\qquad(\boldsymbol{r}\in S_{i}).
\end{alignat}
The chemohydrodynamic problem is to determine the Cauchy stress in
the bulk and the traction at the boundaries in terms of the prescribed
active flux $j^{\mathcal{A}}$. The linearity of the diffusion and
Stokes equations and their respective boundary conditions, together
with the linearity of slip flow equation that couples them, makes
it possible to obtain a formally exact solution to the chemohydrodynamic
problem. We show this below. 

First, we take advantage of the spherical symmetry of the colloids
to parametrize the fields on their surfaces in terms of the $l$-th
rank irreducible tensorial harmonics $\mathbf{Y}^{(l)}(\bm{\hat{\rho}})=(-1)^{l}\rho^{l+1}\boldsymbol{\nabla}^{l}\rho^{-1}$.
The expansion of the surface concentration and surface flux are\begin{subequations}\label{eq:concGalerkin}
\begin{alignat}{1}
c(\boldsymbol{\rho}_{i}) & =\sum_{m=0}^{\infty}w_{m}\mathbf{C}_{i}^{(m)}\cdot\mathbf{Y}^{(m)}(\bm{\hat{\rho}}_{i}),\\
j^{\mathcal{A}}(\boldsymbol{\rho}_{i}) & =\sum_{m=0}^{\infty}\tilde{w}_{m}\mathbf{J}_{i}^{(m)}\cdot\mathbf{Y}^{(m)}(\bm{\hat{\rho}}_{i}),\quad(\text{specified})
\end{alignat}
\end{subequations}where $\mathbf{C}_{i}^{(l)}$ and $\mathbf{J}_{i}^{(l)}$
are $l$-th rank symmetric irreducible tensorial coefficients with
$(2l+1)$ independent components \cite{hess2015tensors}. Here and
below, a maximal contraction of two tensors is denoted by a dot product.
The coefficients $\mathbf{J}_{i}^{(m)}$ are specified as part of
the problem and, for uniaxially symmetric activity, can be be parametrized
as $\mathbf{J}_{i}^{(m)}=J^{(m)}\mathbf{Y}^{(m)}(\bm{p}_{i})$, reducing
the number of free parameters considerably. The first four uniaxial
tensorial surface modes are shown in Fig.(\ref{fig:1chem}). The corresponding
expansions for the active slip and the traction are \begin{subequations}\label{slipTractionExpansion}
\begin{align}
\boldsymbol{v}^{\mathcal{A}}(\boldsymbol{\rho}_{i}) & =\sum_{l=1}^{\infty}w_{l-1}\mathbf{V}_{i}^{(l)}\cdot\mathbf{Y}^{(l-1)}(\bm{\hat{\rho}}_{i}),\quad\\
\boldsymbol{f}^{\mathcal{A}}(\boldsymbol{\rho}_{i}) & =\sum_{l=1}^{\infty}\tilde{w}_{l-1}\mathbf{F}_{i}^{(l)}\cdot\mathbf{Y}^{(l-1)}(\bm{\hat{\rho}}_{i}),\quad(\text{sought})
\end{align}
\end{subequations}where $\mathbf{V}_{i}^{(l)}$ and $\mathbf{F}_{i}^{(l)}$
are $l$-th rank tensorial coefficients, symmetric and irreducible
in their last $l-1$ indices, with the dimensions of force and velocity
respectively. The $l$-dependent expansion weights are
\begin{equation}
w_{l}=\frac{1}{l!(2l-1)!!},\qquad\tilde{w}_{l}=\frac{2l+1}{4\pi b^{2}},
\end{equation}
and unless otherwise specified, sums over repeated indices $m,m'$
start from $0$ while sums over repeated indices $l,l'$ start from
$1$. 

Second, linearity of the diffusion equation and the boundary conditions
implies that the tensorial coefficients of the surface concentration
and surface flux are linearly related,
\begin{equation}
\mathbf{C}_{i}^{(m)}=-\boldsymbol{\varepsilon}{}_{ik}^{(m,m')}\cdot\mathbf{J}_{k}^{(m')}.\label{eq:elastance}
\end{equation}
In the above, repeated harmonic ($m,m'$) and particle ($i,k$) indices
are summed over. The coefficients of proportionality, $\boldsymbol{\varepsilon}{}_{ik}^{(m,m')}$,
are tensors of rank $(m+m')$ and many-body functions of the particle
positions. Maxwell, in his study of the capacitance of a system of
spherical conductors, called the coefficients $\boldsymbol{\varepsilon}{}_{ik}^{(0,0)}$
elastances \cite{maxwell1881treatise}. We call the complete set of
coefficients generalized elastance tensors.

Third, linearity of the slip equation implies that the tensorial coefficients
of the slip and surface concentration are linearly related,
\begin{alignat}{1}
\mathbf{V}_{i}^{(l)}=- & \boldsymbol{\chi}^{(l,m)}\cdot\mathbf{C}_{i}^{(m)},\label{eq:slipls-1}
\end{alignat}
where $\boldsymbol{\chi}^{(l,m)}$ is a coupling tensor of rank $(l+m)$
that depends on the phoretic mobility $\mu_{c}.$ We assume the latter
to not vary between particles, making the coupling tensors independent
of the particle indices. 

Fourth, linearity of the Stokes equation and the boundary conditions
implies that the tensorial coefficients of the slip and traction are
linearly related,
\begin{eqnarray}
\mathbf{F}_{i}^{(l)} & = & -\boldsymbol{\gamma}_{ik}^{(l,l')}\cdot\mathbf{V}_{k}^{(l')},\label{eq:generalizedStokesLaws}
\end{eqnarray}
where the coefficients of proportionality, $\boldsymbol{\gamma}_{ik}^{(l,l')}$,
are tensors of rank $(l+l')$ and many-body functions of the particle
positions. We call the complete set of coefficients the generalized
friction tensors \cite{singh2018generalized}. 

Finally, eliminating the coefficients of the concentration and slip
between the preceding three equations, we obtain a direct relation
between the prescribed coefficients of the active flux and the sought
coefficients of the traction,
\begin{equation}
\mathbf{F}_{i}^{(l)}=-\underbracket[0.6pt]{\boldsymbol{\gamma}_{ij}^{(l,l')}}_{\text{Stokes}}\cdot\boldsymbol{\chi}^{(l',m')}\cdot\underbracket[0.6pt]{\boldsymbol{\varepsilon}{}_{jk}^{(m',m)}}_{\text{Laplace}}\cdot\mathbf{J}{}_{k}^{(m)}.\label{eq:activeTraction}
\end{equation}
This shows that the force per unit area on autophoretic colloids has
both many-body chemical and hydrodynamic contributions encoded, respectively,
in the generalized elastance and friction tensors and determined,
respectively, by solutions of the Laplace and Stokes equations. Ignoring
chemical (hydrodynamic) interactions between particles amounts to
setting the components of the elastance (friction) tensors off-diagonal
in the \emph{particle} indices to zero. The coefficients $\mathbf{F}_{i}^{(1)}$
and $b\boldsymbol{\,\epsilon}\cdot\mathbf{F}_{i}^{(2)}$, where $\boldsymbol{\epsilon}$
is the Levi-Civita tensor, are the active force and torque on the
$i$-th colloid that determine its rigid body motion. The remaining
coefficients are required to determine suspension-scale quantities
such as the rheological response and the power dissipation \cite{singh2018generalized}.
The formal solution above is completed by providing expressions for
the elastance, coupling and friction tensors. 

The coupling tensors are obtained straightforwardly from the slip
flow equation Eq.(\ref{eq:slipls-1}) with a prescribed mobility,
\begin{equation}
\boldsymbol{\chi}^{(l,m)}=-\int\tilde{w}_{l-1}\mathbf{Y}^{(l-1)}(\hat{\bm{\rho}}_{i})\mu_{c}(\boldsymbol{\rho}_{i})w_{m}\boldsymbol{\nabla}_{s}\mathbf{Y}^{(m)}(\hat{\bm{\rho}}_{i})\,dS_{i}.\label{eq:coupling}
\end{equation}
The elastance tensors are most conveniently obtained from the solution
of the boundary integral representation of the Laplace equation. To
leading order in distance, they assume a pairwise form given in terms
of gradients of the Green's function $H$ of the Laplace equation,
\[
-\boldsymbol{\varepsilon}_{ik}^{(m,m')}\approx b^{m+m'}\bm{\nabla}_{{\scriptscriptstyle \boldsymbol{R}_{i}}}^{m}\bm{\nabla}_{{\scriptscriptstyle \boldsymbol{R}_{k}}}^{m'}H(\boldsymbol{R}_{i},\boldsymbol{R}_{k}).
\]
A method for calculating their many-body forms, to any desired order
of accuracy, is provided in the Appendix \ref{app:solutionBoundaryIntegrals}.
The generalized friction tensors are similarly obtained from the boundary
integral representation of the Stokes equation \cite{singh2018generalized}.
To leading order in distance, they assume a pairwise form given in
terms of gradients of the Green's function $\mathbf{G}$ of the Stokes
equation
\[
-\boldsymbol{\gamma}_{ik}^{(l,\,l^{\prime})}\approx b^{l+l'-2}\bm{\nabla}_{{\scriptscriptstyle \boldsymbol{R}_{i}}}^{l-1}\bm{\nabla}_{{\scriptscriptstyle \boldsymbol{R}_{k}}}^{l'-1}\mathbf{G}(\boldsymbol{R}_{i},\boldsymbol{R}_{k}).
\]
A method for calculating them, to any desired order of accuracy, has
been presented earlier \cite{singh2018generalized}. Combining the
three preceding equations with Eq.(\ref{eq:activeTraction}) gives
explicit expressions for the active traction in terms of the particle
positions, orientations and activities. The Cauchy stress in the fluid
is then determined by the integral representation that relates it
to the traction \cite{ghose2014irreducible,singh2018generalized}.

We now derive the dynamical equations for autophoretic motion of the
colloids from the balance of forces and torques. The active force
$\mathbf{F}_{i}^{\mathcal{A}}$ and torque $\mathbf{T}_{i}^{\mathcal{A}}$
on the $i$-th colloid follows from the first two modes of the active
traction in (\ref{eq:activeTraction}):\begin{subequations}\label{eq:forceActive}
\begin{align}
\mathbf{F}_{i}^{\mathcal{A}} & =-\boldsymbol{\gamma}_{ij}^{(T,l)}\cdot\boldsymbol{\chi}^{(l,m')}\cdot\boldsymbol{\varepsilon}{}_{jk}^{(m',m)}\cdot\mathbf{J}{}_{k}^{(m)},\\
\mathbf{T}_{i}^{\mathcal{A}} & =-\boldsymbol{\gamma}_{ij}^{(R,l)}\cdot\boldsymbol{\chi}^{(l,m')}\cdot\boldsymbol{\varepsilon}{}_{jk}^{(m',m)}\cdot\mathbf{J}{}_{k}^{(m)}.
\end{align}
\end{subequations} where $\boldsymbol{\gamma}_{ij}^{(T,l)}=\boldsymbol{\gamma}_{ij}^{(1,l)}$
and $\boldsymbol{\gamma}_{ij}^{(R,l)}=\boldsymbol{\epsilon}\cdot\boldsymbol{\gamma}_{ij}^{(2,l)}$.
The drag forces and torques are given by the standard expressions
\cite{mazur1982,ladd1988}\begin{subequations}
\begin{align}
\mathbf{F}_{i}^{\mathcal{D}} & =-\boldsymbol{\gamma}_{ik}^{TT}\cdot\mathbf{V}_{k}-\boldsymbol{\gamma}_{ik}^{TR}\cdot\boldsymbol{\Omega}_{k},\\
\mathbf{T}_{i}^{\mathcal{D}} & =-\boldsymbol{\gamma}_{ik}^{RT}\cdot\mathbf{V}_{k}-\boldsymbol{\gamma}_{ik}^{RR}\cdot\boldsymbol{\Omega}_{k}.
\end{align}
\end{subequations} Conservative body forces and torques, $\mathbf{F}_{i}^{\mathcal{P}}$
and $\mathbf{T}_{i}^{\mathcal{P}}$, may, in addition, act on the
particles. Then, Newton's equations for the $i$-th colloid are\begin{subequations}\label{eq:forceBalance}
\begin{align}
M\dot{\mathbf{V}}_{i}=\mathbf{F}_{i}^{\mathcal{A}}+\mathbf{F}_{i}^{\mathcal{D}} & +\mathbf{F}_{i}^{\mathcal{P}},\\
I\dot{\mathbf{\Omega}}_{i}=\mathbf{T}_{i}^{\mathcal{A}}+\mathbf{T}_{i}^{\mathcal{D}} & +\mathbf{T}_{i}^{\mathcal{P}}.
\end{align}
\end{subequations}where $M$ is the mass and $I$ the moment of inertia.
At the colloidal scale, inertia can be ignored and Newton equations
reduced to implicit algebraic equations for the velocities and angular
velocities. Explicit solutions can be obtained in terms of the active
forces and torques (see Appendix \ref{app:Simulation-Details}) and
used to evolve the position and orientation according to the kinematic
equations
\begin{equation}
\dot{\boldsymbol{R}}_{i}=\mathbf{V}_{i},\qquad\dot{\boldsymbol{p}}_{i}=\mathbf{\Omega}_{i}\times\boldsymbol{p}_{i}.
\end{equation}
Autophoretic colloidal motion is thus fully determined by the solution
of the chemohydrodynamic problem. 

\section{Autophoresis of bottom-heavy colloids near a plane wall\label{sec:minimalModel}}

We now apply the above general solution to a specific, experimentally
relevant, situation: autophoretic motion of particles near a planar
no-flux, no-slip wall. Motivated by a minimal representation of autophoretic
Janus colloids, we assume that the active surface flux $j^{\mathcal{A}}$
has monopolar and dipolar modes and that the phoretic mobility is
constant,
\begin{figure}[t]
\centering\includegraphics[width=0.46\textwidth]{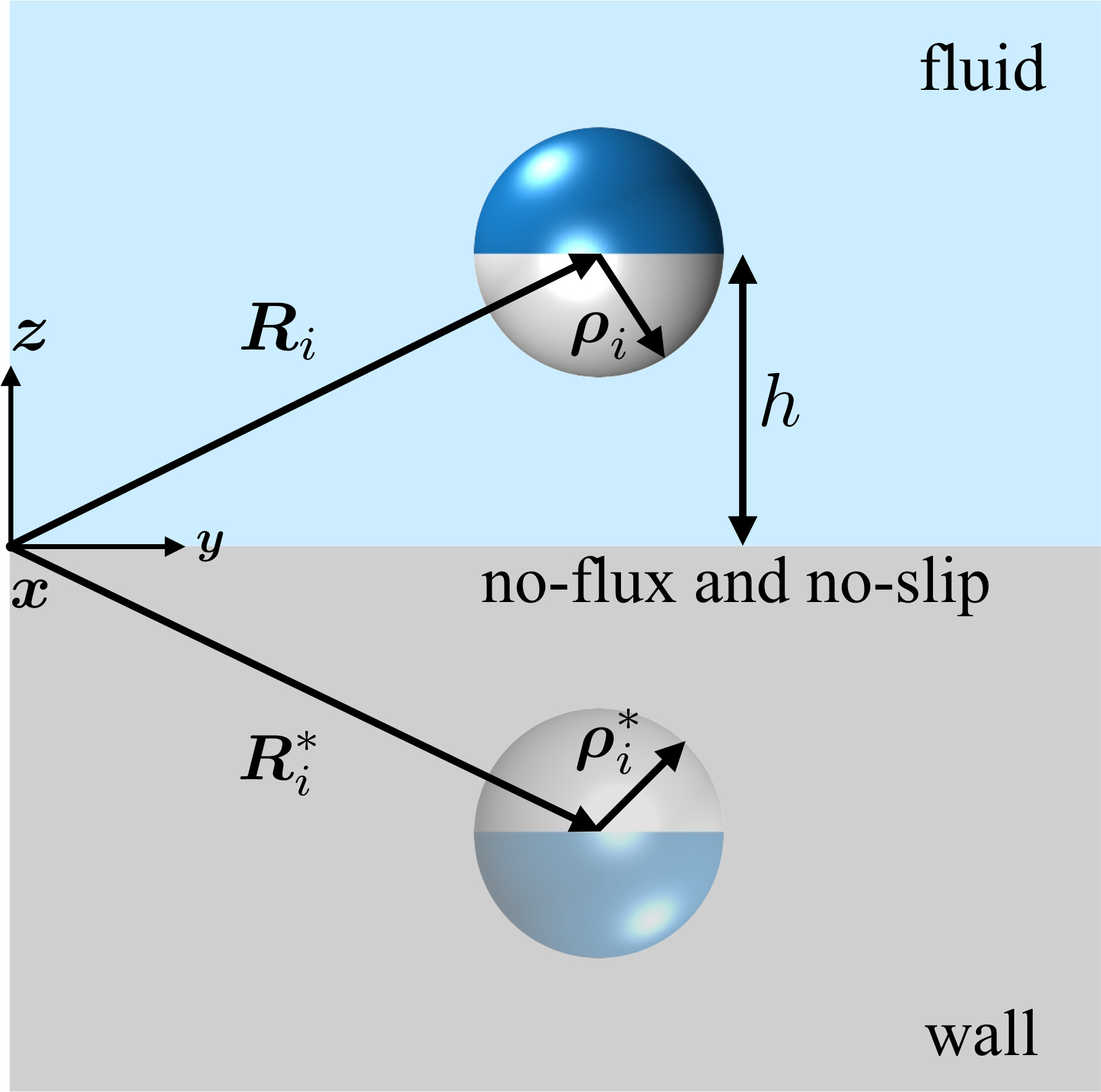}\caption{The system of coordinates used to describe active colloids above a
planar wall. The height, $h$, of the colloid from the wall is explicitly
indicated, and the system of coordinates of its image centered at
$\boldsymbol{R}_{i}^{*}$ below the wall is also shown (see text).
Note that in our method the fluid extends to infinity in the horizontal
and upwards directions; we have no need for periodic or other boundary
conditions at the edges of the simulation box. Indeed, there is no
simulation box, just a set of particle coordinates which can in principle
be anywhere in the upper half space. \label{fig:1corrd-1}}
\end{figure}
\begin{equation}
4\pi b^{2}\,j^{\mathcal{A}}(\boldsymbol{\rho}_{i})=J^{(0)}+3\,J^{(1)}\boldsymbol{p}_{i}\cdot\boldsymbol{\hat{\rho}}_{i},\qquad4\pi b^{2}\,\mu_{c}(\boldsymbol{\rho}_{i})=M^{(0)}.\label{eq:minimal}
\end{equation}
This model has three independent parameters. The assumption of constant
mobility leads to a coupling tensor with exactly one non-zero component,
\begin{gather}
\boldsymbol{\chi}^{(l,m)}=-\delta_{l1}\delta_{m1}\frac{M^{(0)}}{4\pi b^{3}}\boldsymbol{I},\label{eq:couplingMinimal}
\end{gather}
corresponding to $l=m=1$. This restricts the sums in Eq.(\ref{eq:forceActive})
for the active forces and torques to terms where the second harmonic
index of the friction tensor and the first harmonic index of the elastance
tensor are both equal to one. With this simplification, the active
force on a colloid located at $\boldsymbol{R}_{1}$ and oriented along
$\boldsymbol{p}_{1}$ contains exactly two terms:
\begin{alignat}{1}
\mathbf{F}_{1}^{\mathcal{\text{\text{self}}}}=-\underbrace{J^{(0)}\boldsymbol{\gamma}_{11}^{(T,1)}\cdot\boldsymbol{\varepsilon}{}_{11}^{(1,0)}}_{\text{wall-induced propulsion}} & -\underbrace{J^{(1)}\boldsymbol{\gamma}_{11}^{(T,1)}\cdot\boldsymbol{\varepsilon}{}_{11}^{(1,1)}\cdot\boldsymbol{p}_{1}}_{\text{polar self-propulsion}},\label{eq:selfF}
\end{alignat}
The first term is the force arising from the broken spherical symmetry
of the chemical monopole field in the vicinity of the wall and the
second term is the self-propulsion force of the chemical dipole. Thus,
an apolar active colloid ($J^{(0)}\ne0,J^{(1)}=0)$ which, by symmetry,
can have no motion in an unbounded medium, will acquire motion near
a wall due the breaking of spherical symmetry. The scaling of the
force by $-M^{(0)}/4\pi b^{3}$ in the above equation and the three
equations below are for ease of display. With the introduction of
a second colloid, located at $\boldsymbol{R}_{2}$ with orientation
$\boldsymbol{p}_{2}$, the active force on particle $1$ will contain
six additional interaction terms involving particle $2$, \begin{subequations}
\begin{alignat}{1}
\mathbf{F}_{1}^{\text{CI}}= & -J^{(0)}\boldsymbol{\gamma}_{11}^{(T,1)}\cdot\boldsymbol{\varepsilon}{}_{12}^{(1,0)}-J^{(1)}\boldsymbol{\gamma}_{11}^{(T,1)}\cdot\boldsymbol{\varepsilon}{}_{12}^{(1,1)}\cdot\boldsymbol{p}_{2},\\
\mathbf{F}_{1}^{\text{HI}}= & -J^{(0)}\boldsymbol{\gamma}_{12}^{(T,1)}\cdot\boldsymbol{\varepsilon}{}_{22}^{(1,0)}-J^{(1)}\boldsymbol{\gamma}_{12}^{(T,1)}\cdot\boldsymbol{\varepsilon}{}_{22}^{(1,1)}\cdot\boldsymbol{p}_{2},\\
\mathbf{F}_{1}^{\text{CHI}}= & -\negmedspace\negmedspace\underbrace{J^{(0)}\boldsymbol{\gamma}_{12}^{(T,1)}\cdot\boldsymbol{\varepsilon}{}_{21}^{(1,0)}}_{\text{orientation-independent}}\negmedspace-\underbrace{J^{(1)}\boldsymbol{\gamma}_{12}^{(T,1)}\cdot\boldsymbol{\varepsilon}{}_{21}^{(1,1)}\cdot\boldsymbol{p}_{1}}_{\text{orientation-dependent}},
\end{alignat}
\end{subequations}giving a total of eight terms. Every such term
is the product of an elastance tensor and a friction tensor, each
of which can be diagonal or off-diagonal in the particle indices,
giving four distinct kinds of interactions. Terms in which both the
elastance and friction are diagonal represent self-propulsion and
self-interaction forces; terms in which the elastance is off-diagonal
and the friction diagonal represent chemical interactions (CI); terms
in which the elastance is diagonal and the friction off-diagonal represent
hydrodynamic interactions (HI); and terms in which both elastance
and friction are off-diagonal represent chemohydrodynamic interactions
(CHI). The CI are due to chemical gradients induced on particle $1$
due to activity of particle $2$, the HI are due to the flow incident
on particle $1$ due to the slip on particle $2$ and the CHI are
due to the flow incident on particle $1$ from the additional slip
on particle $2$ from the chemical gradient induced on it by the activity
of particle $1$. The monopolar and dipolar modes of activity yield
forces that are, respectively, orientation-independent and orientation-dependent.
The force on particle $2$ is obtained by interchanging indices. Structurally
similar expressions are obtained for the active torques. 

Explicit expressions for these forces and torques are obtained from
the solutions of the linear systems defining the elastance and friction
tensors, as explained in \ref{app:solutionBoundaryIntegrals}. The
coefficients of these linear systems are determined by the Green's
functions of the Laplace and Stokes equations, corresponding to the
boundary conditions imposed, respectively, on the chemical and flow
fields at the boundaries. Here we use solutions obtained at the second
step of a Jacobi iteration, which corresponds to evaluating elastance
and friction tensors to next-to-leading order in distance. We emphasize
that this iteration can be continued to as many steps as needed for
a prescribed accuracy. The linear system may also be solved by direct
methods but we do not pursue this here. The no-flux Green's function
of the Laplace equation is
\begin{equation}
H^{\text{w}}(\boldsymbol{R}_{i},\boldsymbol{R}_{k})=H^{0}(\boldsymbol{r})+H^{0}(\boldsymbol{r}^{*}).\label{eq:laplaceGwall}
\end{equation}
Here $8\pi DH^{0}(\boldsymbol{r})=\nabla^{2}r$ is the Green's function
in an unbounded domain, $\boldsymbol{r}=\boldsymbol{R}_{i}-\boldsymbol{R}_{k}$,
$\boldsymbol{r}^{*}=\boldsymbol{R}_{i}-\boldsymbol{R}_{k}^{*}$, $\boldsymbol{R}_{k}^{*}=\boldsymbol{\mathcal{M}\cdot}\boldsymbol{R}_{k}$,
and $\boldsymbol{\mathcal{M}}=\boldsymbol{I}-2\mathbf{\hat{z}}\mathbf{\hat{z}}$
is the mirror operator with respect to the wall at $z=0$. See Fig.(\ref{fig:1corrd-1})
for the system of coordinates. From Appendix (\ref{app:solutionBoundaryIntegrals}),
the leading forms of elastance tensors relevant to our minimal model
are,
\begin{gather}
\boldsymbol{\varepsilon}{}_{11}^{(1,0)}=\frac{b\hat{\boldsymbol{z}}}{16\pi Dh^{2}},\quad\boldsymbol{\varepsilon}{}_{11}^{(1,1)}=\frac{1}{4\pi bD}+\frac{b^{2}}{4\pi D}\frac{\boldsymbol{I}-3\boldsymbol{\hat{z}}\boldsymbol{\hat{z}}}{8h^{3}},\label{eq:force2Body-1}\\
\boldsymbol{\varepsilon}{}_{12}^{(1,0)}=\frac{b}{4\pi D}\left(\frac{\hat{\boldsymbol{r}}}{r^{2}}+\frac{\hat{\boldsymbol{r}}^{*}}{r^{*2}}\right),\quad\boldsymbol{\varepsilon}{}_{12}^{(1,1)}=\frac{b^{2}}{4\pi D}\left(\frac{\boldsymbol{I}-3\boldsymbol{\hat{r}}\boldsymbol{\hat{r}}}{r^{3}}+\frac{\boldsymbol{I}-3\boldsymbol{\hat{r}^{*}}\boldsymbol{\hat{r}^{*}}}{r^{*3}}\right).
\end{gather}
The no-slip Green's function of Stokes equation is the Lorentz-Blake
tensor \cite{lorentz1896eene,blake1971c}\begin{subequations}
\begin{gather}
G_{\alpha\beta}^{\text{w}}(\boldsymbol{R}_{i},\boldsymbol{R}_{k})=G_{\alpha\beta}^{0}(\boldsymbol{r})+G_{\alpha\beta}^{*}(\boldsymbol{r}^{*}),\label{eq:stokesGwall}\\
G_{\alpha\beta}^{*}(\boldsymbol{r}^{*})=G_{\alpha\beta}^{0}(\boldsymbol{r}^{*})-\big[2h\nabla_{{\scriptscriptstyle \mathbf{r}^{*}}}G_{\alpha3}^{0}(\mathbf{r}^{*})-h^{2}\nabla_{{\scriptscriptstyle \mathbf{r}^{*}}}^{2}G_{\alpha\gamma}^{0}(\mathbf{r}^{*})\big]\mathcal{M}_{\beta\gamma},
\end{gather}
\end{subequations}where $8\pi\eta G_{\alpha\beta}^{0}(\boldsymbol{r})=\left(\nabla^{2}\delta_{\alpha\beta}-\nabla_{\alpha}\nabla_{\beta}\right)r$,
the Oseen tensor, is the Green's function in an unbounded domain and
$G_{\alpha\beta}^{*}(\boldsymbol{r}^{*})$ is the correction required
to satisfy the boundary condition on the plane wall. From \cite{singh2018generalized},
the leading form of the relevant friction tensors are
\begin{equation}
\boldsymbol{\gamma}_{11}^{(T,1)}=\left(\begin{array}{ccc}
\gamma_{{\scriptscriptstyle \parallel}} & 0 & 0\\
0 & \gamma_{{\scriptscriptstyle \parallel}} & 0\\
0 & 0 & \gamma_{\perp}
\end{array}\right),\qquad\boldsymbol{\gamma}_{12}^{(T,1)}=-\left(\begin{array}{ccc}
\gamma_{{\scriptscriptstyle \parallel}}\gamma_{{\scriptscriptstyle \parallel}}G_{xx}^{\text{w}} & \gamma_{{\scriptscriptstyle \parallel}}\gamma_{{\scriptscriptstyle \parallel}}G_{xy}^{\text{w}} & \gamma_{{\scriptscriptstyle \parallel}}\gamma_{\perp}G_{xz}^{\text{w}}\\
\gamma_{{\scriptscriptstyle \parallel}}\gamma_{{\scriptscriptstyle \parallel}}G_{yx}^{\text{w}} & \gamma_{{\scriptscriptstyle \parallel}}\gamma_{{\scriptscriptstyle \parallel}}G_{yy}^{\text{w}} & \gamma_{{\scriptscriptstyle \parallel}}\gamma_{\perp}G_{yz}^{\text{w}}\\
\gamma_{\perp}\gamma_{{\scriptscriptstyle \parallel}}G_{zx}^{\text{w}} & \gamma_{\perp}\gamma_{{\scriptscriptstyle \parallel}}G_{zy}^{\text{w}} & \gamma_{\perp}\gamma_{\perp}G_{zy}^{\text{w}}
\end{array}\right).
\end{equation}
Here $\gamma_{{\scriptscriptstyle \parallel}}=6\pi\eta b(1-6\pi\eta bG_{xx}^{\text{*}})$
and $\gamma_{\perp}=6\pi\eta b(1-6\pi\eta bG_{zz}^{*})$ are the friction
coefficients of a colloid at a height $h$ in the directions parallel
($\parallel$) and perpendicular ($\perp$) to the wall \cite{kim2005}.
\begin{figure*}
\centering\includegraphics[width=0.94\textwidth]{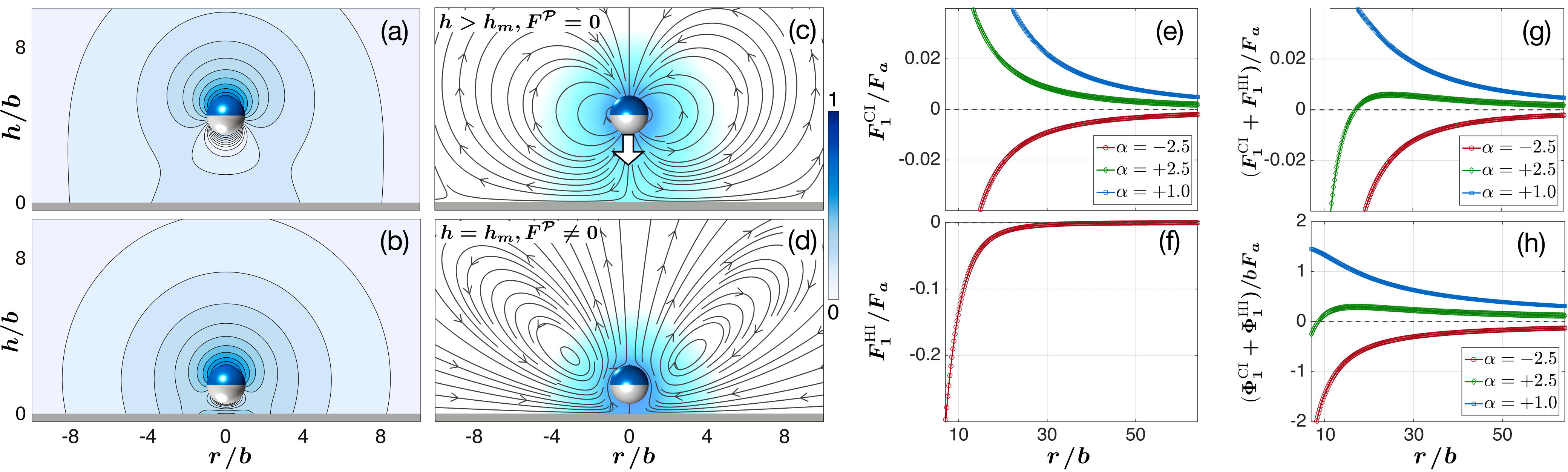}\caption{One and two-body dynamics of autophoretic colloids near a plane wall.
Panel (a-b) are contour-maps of the chemical field around a colloid
at two heights from the wall, while (c-d) are the corresponding flow
streamlines overlaid on the normalized logarithm of the flow speed.
Panel (e) shows the quantitative variation of planar chemical forces,
while panel (f) contains the attractive hydrodynamic force and panel
(g) is the sum of the forces. The corresponding autophoretic non-equilibrium
potential is plotted in panel (h) for three values of dimensionless
number $\alpha$. Panel (a-d) are for $\alpha=2.5$. Here $F_{a}=6\pi\eta b\,v_{s}$.\label{fig:1}}
\end{figure*}

Using the above in Eq.(\ref{eq:selfF}) and retaining leading terms
in distance gives the self-contribution to the active force to be
\begin{alignat}{1}
\mathbf{F}_{1}^{\mathcal{\text{self}}}=\, & \gamma_{{\scriptscriptstyle \perp}}v_{s}\left(\boldsymbol{p}_{1}+\frac{b^{2}}{2\alpha h^{2}}\hat{\boldsymbol{z}}\right).\label{eq:selfF-1}
\end{alignat}
where 
\[
v_{s}=\frac{M^{(0)}J^{(\text{1})}}{16\pi^{2}b^{4}D},\qquad\alpha=\frac{J^{(1)}}{J^{(0)}},
\]
are, respectively, the intrinsic self-propulsion velocity and ratio
of dipolar to monopolar activity. The propulsive force is directed
along or opposite to the orientation accordingly as the product $M^{(0)}J^{(1)}$
is positive or negative. For a given sign of $v_{s}$ the interaction
with the wall depends only on the activity ratio. For a particle with
$v_{s},\alpha>0$, forces are balanced at a height $h^{2}=b^{2}/2\alpha$,
and a self-levitating state results. For other sign combinations,
the particle is either repelled from or attracted to the wall. In
the latter case, when the particle is brought to rest by steric interactions
at a height $h_{m}$, the Stokeslet (force monopole on the colloid)
on it points normal to and away from the wall. The chemical and flow
fields produced by a particle at heights $h>h_{m}$ and $h=h_{m}$
are shown in Fig.( \ref{fig:1}). 

The CI force to leading order is 
\begin{alignat}{1}
\mathbf{F}_{1}^{\text{CI}}(\boldsymbol{r},\boldsymbol{p}_{2}) & =2\gamma_{{\scriptscriptstyle \parallel}}v_{s}\left(\frac{b^{2}\hat{\boldsymbol{r}}}{\alpha r^{2}}+\frac{b^{3}}{r^{3}}\left[\boldsymbol{p}_{2}-3(\boldsymbol{p}_{2}\cdot\hat{\boldsymbol{r}})\hat{\boldsymbol{r}}\right]+\frac{b^{2}\hat{\boldsymbol{r}}^{*}}{\alpha r^{*2}}+\frac{b^{3}}{r^{*3}}\left[\boldsymbol{p}_{2}-3(\boldsymbol{p}_{2}\cdot\hat{\boldsymbol{r}}^{*})\hat{\boldsymbol{r}}^{*}\right]\right),\label{eq:chemforce2Body}
\end{alignat}
with orientation-independent inverse-square and orientation-dependent
inverse-cube contributions. From the expression for the force on particle
$2$, obtained by interchanging indices, it is clear that the forces
are non-reciprocal and violate the action-reaction principle. At a
fixed height, the CI force can be expressed as the gradient, with
respect to in-plane coordinates, of a non-equilibrium potential,
\begin{equation}
\Phi_{\text{1}}^{\text{CI}}(\boldsymbol{r},\boldsymbol{p}_{2})=2\gamma_{{\scriptscriptstyle \parallel}}bv_{s}\left(\frac{b}{\alpha r}+\frac{b}{\alpha r^{*}}-\frac{b^{2}}{r^{2}}\left[\boldsymbol{p}_{2}\cdot\hat{\boldsymbol{r}}\right]-\frac{b^{2}}{r^{*2}}\left[\boldsymbol{p}_{2}\cdot\hat{\boldsymbol{r}}^{*}\right]\right).\quad(\text{fixed height)}
\end{equation}
In general, CI terms at any order can be so expressed if the friction
tensors are independent of configuration. Notably, the autophoretic
non-equilibrium potential is orientation-dependent, unlike the orientation-independent
non-equilibrium potentials that appear in phoretic phenomena in externally
imposed gradients \cite{squires2001effective,di2009colloidal}. 

The HI force has a more complicated functional form that simplifies
when particle $2$ is oriented normal to the wall:
\begin{alignat}{1}
\mathbf{F}_{1}^{\text{HI}}(\boldsymbol{r}) & =-\frac{\gamma_{{\scriptscriptstyle \parallel}}\gamma_{\perp}v_{s}}{2\pi\eta}\left(1+\frac{b^{2}}{2\alpha h^{2}}\right)\frac{3h^{3}}{(r^{2}+4h^{2})^{5/2}}\boldsymbol{r}.\label{eq:force2Body-2}
\end{alignat}
It has an explicit dependence on the height of the particle pair from
the wall. With the orientation normal to the wall, the HI force can
also be expressed as the gradient, with respect to in-plane coordinates,
of a non-equilibrium potential, 
\begin{equation}
\Phi_{1}^{\text{HI}}(\boldsymbol{r})=-\frac{\gamma_{{\scriptscriptstyle \parallel}}\gamma_{\perp}v_{s}}{2\pi\eta}\left(1+\frac{b^{2}}{2\alpha h^{2}}\right)\frac{h^{3}}{(r^{2}+4h^{2})^{3/2}},\quad(\boldsymbol{p}_{2}\perp\text{wall}).
\end{equation}
This follows from the irrotational character the flow assumes when
particle $2$ is oriented normal to the wall. In contrast, flow in
a Hele-Shaw geometry (parallel walls separated by a gap small compared
to their size) is irrotational for any orientation \cite{liron1976stokes,thutupalli2018FIPS,kanso2019phoretic}
and non-equilibrium hydrodynamic potentials then exist for arbitrary
orientations. In general, such potentials will fail to express the
HI when the flow contains significant amounts of vorticity. The CHI
contains product of gradients of the Laplace and Stokes Green's functions
and, therefore, is never the gradient of a potential. However, as
it has a weaker distance-dependence than both CI and HI, the potentials,
when they exist, capture the leading contributions to the forces. 

We plot our results for forces and potential in panels (e-h) of Fig.(\ref{fig:1}),
as a function lateral pair separation $r$, at different values of
$\alpha$. It should be noted that the planar chemical force, obtained
from Eq.(\ref{eq:chemforce2Body}), contains contributions up to $O(1/r^{4})$
when it is evaluated for orientation vectors along $\hat{z}$ direction.
In this setting, the next term of the chemical force contributes only
at $O(1/r^{6})$, and is thus, not included for this analysis. In
panel (e), we show that the chemical force $\mathbf{F}_{1}^{\text{CI}}$
depends on the sign of $\alpha$, which is controlled by the sign
of $J^{(0)}$, while the hydrodynamic component $\mathbf{F}_{1}^{\text{HI}}$
is always attractive \cite{singh2016crystallization}. Thus, the effective
potential has a barrier if the chemical interaction is repulsive,
as shown in panel (h); the effective interaction is then repulsive
for $r>r_{c}$, and attractive for $r<r_{c}$ where
\begin{alignat}{1}
r_{c} & \sim h\sqrt{\frac{3\alpha h\gamma_{\perp}}{4\pi b^{2}\eta}}.\label{eq:clusterSize}
\end{alignat}
is an interaction scale that depends on the height from the wall.
The emergence of a length scale in slow viscous flow near a wall has
also been noted in another context \cite{driscoll2017unstable}. 

We emphasize conditions in which non-equilibrium potentials exist.
The CI always admits a potential when the friction tensors are constant;
the HI always admits a potential when the flow is irrotational; the
CHI, in general, does not admit a potential. However, as CHI is sub-dominant
in comparison to CI and HI, it is not surprising that the approximation
of forces by potentials leads to good agreement with simulations below,
where no such approximation is made. We do not present a similarly
detailed characterization of the active torque but turn, instead,
to simulations of many-body effects where such torques are included. 

\begin{figure*}[t]
\centering\includegraphics[width=0.98\textwidth]{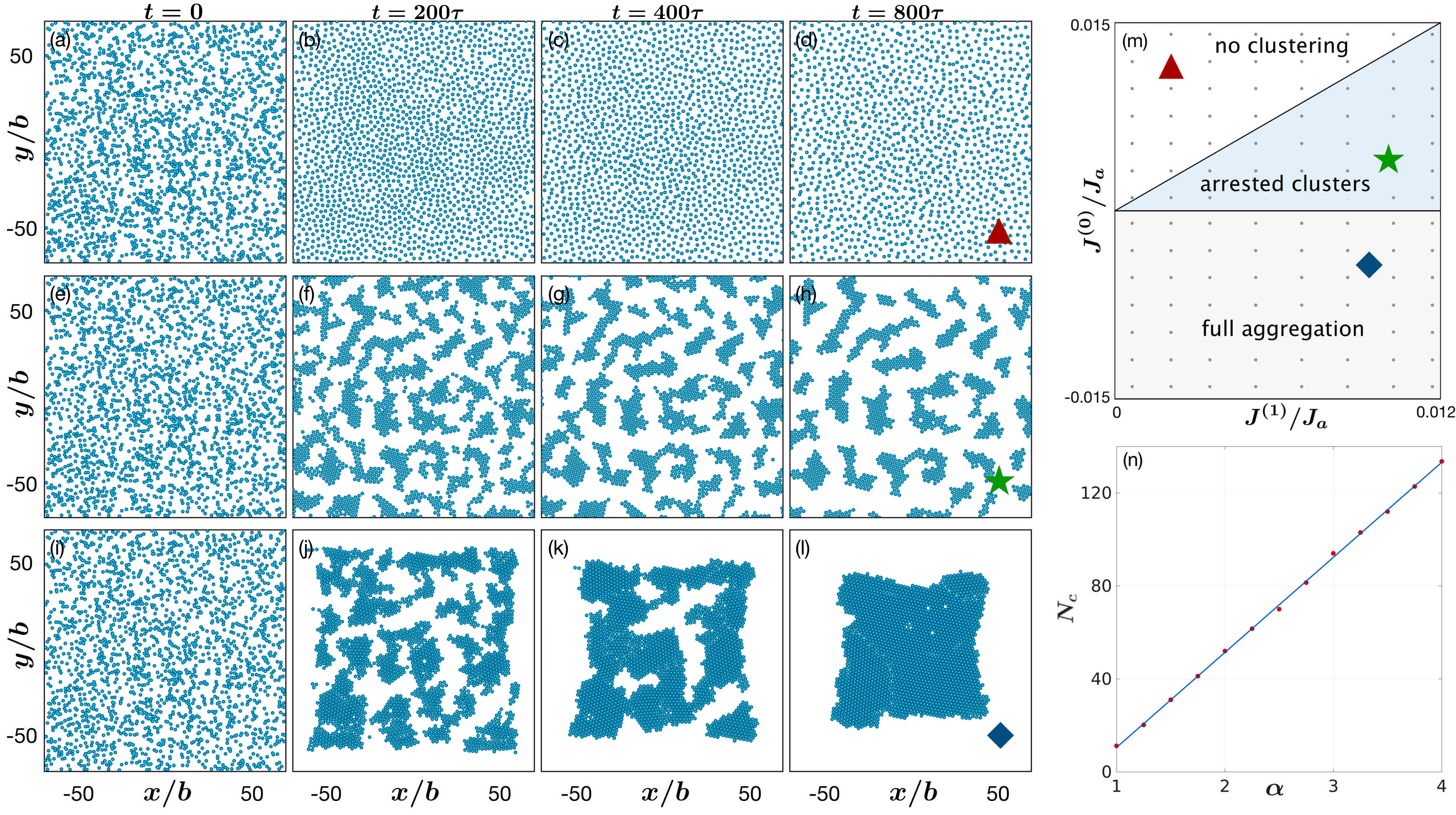}\caption{Self-assembly of $2^{11}$ autophoretic colloids. Panels (a-l) corresponds
to three distinct states as indicated in the state diagram of panel
(m), in the plane of strengths of the chemical dipole $J^{(1)}$ and
monopole $J^{(0)}$ (see (\ref{eq:minimal}) for the model used).
The gray dots denote simulation points, and $J_{a}=Db^{-2}$. Panel
(n) is average number of particles in a cluster ($N_{c}$) as function
of the ratio $\alpha$. Here $\tau=b/v_{s}$. We emphasize that no
periodic boundary condition is used and the system is only confined
by a plane infinite wall. The panels (a-l) have been cropped in the
same extent of space to clearly show the difference in the three regimes.\label{fig:2}}
\end{figure*}

In panel (a-l) of Fig.(\ref{fig:2}), we show the dynamics of $2^{11}$
autophoretic colloids for three different values of the activity number
$\alpha$. In all cases, we start from an initially random hard-sphere
configuration \cite{skoge2006packing}. Panels (a-d) of Fig.(\ref{fig:2})
correspond to $\alpha<1$; here the chemical repulsion dominates the
hydrodynamic attraction. The result is a liquid-like steady-state
as the effective interaction between the colloids is fully repulsive.
For a larger value of $\alpha=2.5$, there is a barrier at $r=r_{c}$
as explained above (see Eq.(\ref{eq:chemforce2Body})). This leads
to arrested phase separation with clusters of particles, as shown
in panels (e-h). The dynamic clusters are similar to those reported
in experiments of autophoretic colloids \cite{palacci2013living,theurkauff2012dynamic}.
When $\alpha$ is negative, both chemical and hydrodynamic interactions
are attractive, and full phase separation is achieved. Thus, we have
identified three distinct phases in the space of chemical parameters;
see Fig.(\ref{fig:2}) (m). We restrict $J^{(1)}$ to positive values
as it corresponds to self-propulsion into the wall, which is necessary
in our model to induce the Stokeslet away from the wall, which leads
to the attractive hydrodynamic forces between the colloids. The chemical
monopole $J^{(0)}$, on the other hand, can take both positive and
negative values. Thus, although the dynamics is determined in terms
of a single ratio $\alpha$, it is useful to show the two-dimensional
phase diagram of Fig.(\ref{fig:2}) (\emph{m}).

The average number of particles $N_{c}$ in the arrested clusters
can be tuned by varying the activity parameter $\alpha$. In panel
(n) of Fig.(\ref{fig:2}), we show that $N_{c}$ is linearly proportional
to $\alpha$. The scaling can be understood from the fact that the
number of particles in a two-dimensional cluster is proportional to
$r_{c}^{2}$, and from Eq.(\ref{eq:clusterSize}) $r_{c}$ is proportional
to $\sqrt{\alpha}$. It is interesting to compare the resulting linear
scaling with the experiments of \cite{theurkauff2012dynamic,buttinoni2013DynamicClustering}.
There, the cluster size grows linearly with the self-propulsion speed
$v_{s}$ of isolated colloids, when this speed is varied by adjusting
the fuel concentration or light intensity. Within our theory, $v_{s}$
is indeed proportional to $\alpha$, but only if the monopole current
$J^{(0)}$ is held fixed as the dipolar activity $J^{(1)}$ is varied.
At present we can see no reason to expect constant $J^{(0)}$ on varying
the overall fuel level, in which case the explanation of the experimental
linear scaling lies beyond the present theory. However, this finding
may offer valuable mechanistic information. Specifically, we assumed
autophoresis to stem from the active surface chemical flux $j^{\mathcal{A}}$
of a single diffusant field $c$, whereas the mechanism of self-propulsion
arising in experimental systems may instead require a description
involving multiple (possibly charged) diffusant species \cite{brown2014ionic,brown2017ionic}.

\section{Discussion\label{sec:disc}}

We have shown that chemical and hydrodynamic many-body effects in
a suspension of autophoretic particles can be fully determined in
terms of elastance, coupling and friction tensors. These tensors can
be calculated from the phoretic mobility and the Green's functions
of the Laplace and Stokes equations that incorporate the appropriate
boundary conditions on the chemical and the flow. 

For particles at fixed heights from, and oriented normal to, a planar
no-flux, no-slip wall, chemical and hydrodynamic forces can be expressed
as in-plane gradients of a non-equilibrium potential. This remarkably
simple pairwise description is fully underpinned by our general treatment
of the many-body coupling between a rapid chemical diffusant, slip
and hydrodynamic interactions, whose numerical investigation in other
geometries we leave to future work. The potential can be purely attractive
or repulsive, or have a barrier. Our barrier heights scale like $F_{a}b\sim6\pi\eta b^{2}\,v_{s}$,
which for a typical experiment \cite{palacci2013living}, with radius
$b=2$ $\mathrm{\mu m}$ and self-propulsion speed $v_{s}=10$ $\mathrm{\mu ms^{-1}}$,
is roughly $10^{-19}$ J, two orders of magnitude higher than the
thermal energy $k_{B}T$. This changes for smaller particles so that
chemical, hydrodynamic and thermal forces are then all relevant. Our
many-body formalism is generalizable to this case, and also, once
they are fully identified, to some of more complex multi-species mechanisms
of autophoresis that may be important experimentally.

\textcolor{black}{Although we have only considered the chemical field,
our formalism is applicable to any harmonic scalar field (for example,
a temperature field \cite{wurger2010thermal}) produced locally by
the colloids or due to an external gradient. In this work, we do not
resolve the near-field hydrodynamic interactions between the colloids.
This can be done in our theory by using lubrication-corrected friction,
mobility, and propulsion tensors \cite{singh2018generalized}. While
our simulations allow for orientational fluctuations, albeit of a
small magnitude, the analytical form of the potential is obtained
in the limit of zero orientational fluctuation. The excellent agreement
between the theory and the simulation confirms that small orientational
fluctuations do not significantly change the form of the potential.
We leave the investigation of non-equilibrium potential, if one exists,
for large orientational fluctuations to future work. In particular,
when the propulsion direction of particles is not constrained to lie
normal to the wall, the tangential component creates the conditions
under which motility-induced phase separation \cite{cates2015} can
be expected to arise. This is a separate mechanism from the one studied
here which can instead be classified as flow-induced phase separation
(FIPS) \cite{singh2016crystallization}. The full problem presumably
combines both of these types of active phase separation in a complex
way. This is one reason why we have focused here on the pure FIPS
limit of strongly aligned particles as the exemplary problem for combining
hydrodynamic and chemical interactions in active colloidal systems.}

\section*{Acknowledgements }

We thank an anonymous referee for constructive remarks on improving
the presentation of our results. RS is funded by a Royal Society-SERB
Newton International Fellowship. RA thanks the Isaac Newton Trust
for an Early Career Support grant. MEC is funded by the Royal Society.
Numerical work was performed on the Fawcett cluster at the Centre
for Mathematical Sciences. Work funded in part by the European Research
Council under the Horizon 2020 Programme, ERC grant agreement number
740269.\begin{appendix}

\section{Boundary integrals solution for elastance tensors\label{app:solutionBoundaryIntegrals}}

The boundary integral equation of the Laplace equation gives the concentration
field on the surface of the $i$-th colloid in terms of and integral
over boundaries of all the colloids
\begin{align}
\tfrac{1}{2}c\,(\boldsymbol{r}_{i})=\int H(\boldsymbol{R}_{i}+\mathbf{\boldsymbol{\rho}}_{i},\,\boldsymbol{R}_{k}+\mathbf{\boldsymbol{\rho}}_{k})\,j^{\mathcal{A}}(\boldsymbol{\rho}_{k})\,\text{d}S_{k}+\int L(\boldsymbol{R}_{i}+\mathbf{\boldsymbol{\rho}}_{i},\,\boldsymbol{R}_{k}+\mathbf{\boldsymbol{\rho}}_{k})\,c(\boldsymbol{\rho}_{k})\,\text{d}S_{k}.\label{eq:bieLaplace}
\end{align}
Here $L=D\boldsymbol{\hat{\rho}}\cdot\mathbf{\boldsymbol{\nabla}}H$
and the integrand vanishes at any other boundary in the bulk fluid
and at the infinity. We use the Galerkin method of solution by expanding
the boundary fields in terms of tensorial spherical harmonics, as
given in Eq.(\ref{eq:concGalerkin}). The principle advantage of the
Galerkin method is that the matrix elements of the linear system,
which are double integrals over the particle boundaries, can be calculated
exactly in terms of the Green's functions, and thus, provides the
greatest accuracy for the least number of discrete degrees of freedom
\cite{muldowney1995spectral,youngren1975stokes}. Such calculations
would be prohibitively expensive in the boundary element method, which
instead collocates the integral equation at specific points on the
boundary.\textcolor{black}{{} }Multiplying both sides of Eq.(\ref{eq:bieLaplace})
by the $m$-th basis function and integrating on the surface of the
\textit{i}-th colloid gives a linear system for the unknown $\mathbf{C}_{i}^{(m)}$
\[
\tfrac{1}{2}\mathbf{C}_{i}^{(m)}=\boldsymbol{H}_{ik}^{(m,\,m')}(\boldsymbol{R}_{i},\boldsymbol{R}_{k})\cdot\mathbf{J}_{k}^{(m')}+\boldsymbol{L}_{ik}^{(m,m')}(\boldsymbol{R}_{i},\boldsymbol{R}_{k})\cdot\mathbf{C}_{k}^{(m')}
\]
Comparing with Eq.(\ref{eq:elastance}), the elastance tensors are
given as 
\begin{gather}
\boldsymbol{\varepsilon}{}_{ik}^{(m,m')}=-\Big[\big(\tfrac{1}{2}\boldsymbol{I}-\boldsymbol{L}\big)^{-1}\boldsymbol{H}\Big]_{ik}^{(m,m')}.\label{eq:elastanceSol}
\end{gather}
Here $\boldsymbol{H}$ and $\boldsymbol{L}$ are matrices, whose $(m,\,m^{\prime})$
element in the $ik$ block are $\boldsymbol{H}_{ik}^{(m,\,m')}(\boldsymbol{R}_{i},\boldsymbol{R}_{k})$
and $\boldsymbol{L}_{ik}^{(m,\,m')}(\boldsymbol{R}_{i},\boldsymbol{R}_{k})$
respectively. The matrix elements of the linear system are\begin{subequations}\label{eq:matrixElements}
\begin{alignat*}{1}
\boldsymbol{H}_{ik}^{(m,m')}(\boldsymbol{R}_{i},\boldsymbol{R}_{k})=\tilde{w}_{m}\tilde{w}_{m'} & \int\mathbf{Y}^{(m)}(\hat{\bm{\rho}}_{i})H(\boldsymbol{R}_{i}+\bm{\rho}_{i},\boldsymbol{R}_{k}+\bm{\rho}_{k})\mathbf{Y}^{(m')}(\hat{\bm{\rho}}_{k})\,d\text{S}_{i}d\text{S}_{k},\\
\boldsymbol{L}_{ik}^{(m,m')}(\boldsymbol{R}_{i},\boldsymbol{R}_{k})=\tilde{w}_{m}w_{m'} & \int\mathbf{Y}^{(m)}(\hat{\bm{\rho}}_{i})L(\boldsymbol{R}_{i}+\bm{\rho}_{i},\boldsymbol{R}_{k}+\bm{\rho}_{k})\mathbf{Y}^{(m')}(\hat{\bm{\rho}}_{k})d\text{S}_{i}d\text{S}_{k}.
\end{alignat*}
\end{subequations}The above integrals are completed by Taylor expansion
of the Green\textquoteright s function and using the orthogonality
of the basis functions ($w_{m}\tilde{w}_{m'}\int\mathbf{Y}^{(m)}(\hat{\bm{\rho}}_{i})\mathbf{Y}^{(m')}(\hat{\bm{\rho}}_{i})\,d\text{S}_{i}=\delta_{mm'}\boldsymbol{\Delta}$)
and standard Bessel integrals to obtain \cite{singh2015many}
\begin{alignat*}{1}
\boldsymbol{H}_{ik}^{(m,m')}(\boldsymbol{R}_{i},\boldsymbol{R}_{k}) & =\begin{cases}
\frac{\delta_{ik}\delta_{mm'}}{4\pi bDw_{m}}\mathbf{\Delta}^{(m)} & {\displaystyle k=i,}\\
{\displaystyle b^{m+m'}\bm{\nabla}_{{\scriptscriptstyle \boldsymbol{R}_{i}}}^{m}\bm{\nabla}_{{\scriptscriptstyle \boldsymbol{R}_{k}}}^{m'}H(\boldsymbol{R}_{i},\boldsymbol{R}_{k})} & {\displaystyle k\neq i,}
\end{cases}\\
\boldsymbol{L}_{ik}^{(m,m')}(\boldsymbol{R}_{i},\boldsymbol{R}_{k}) & =\begin{cases}
-\frac{1}{2}\mathbf{\Delta}^{(m)}\qquad & {\displaystyle k=i,}\\
e_{m}\bm{\nabla}_{{\scriptscriptstyle \boldsymbol{R}_{i}}}^{m}\bm{\nabla}_{{\scriptscriptstyle \boldsymbol{R}_{k}}}^{m'}H(\boldsymbol{R}_{i},\boldsymbol{R}_{k})\quad\,\, & {\displaystyle k\neq i.}
\end{cases}
\end{alignat*}
Here $e_{m}=w_{m-1}Db^{(m+m'-5)}/(4\pi\tilde{w}_{m}\tilde{w}_{m-1})$
and $\mathbf{\Delta}^{(m)}$ is a tensor of rank $2m$, which reduces
a tensor of rank $m$ to its symmetric irreducible form. 

With the matrix elements so determined, the linear system can be solved
by a variety of methods. Here, we use the venerable ``method of reflections''
due to Smoluchowski, which is the Jacobi iteration in disguise. The
solution after the $n$-th iteration is \cite{saad2003iterative}
\begin{alignat}{1}
\Big(\boldsymbol{\varepsilon}_{ik}^{(m,m')}\Big)^{[n]} & =-\frac{1}{\mathit{A}_{ii}^{(m,m)}}\bigg[\boldsymbol{H}_{ik}^{(m,m')}-\boldsymbol{\sum^{\prime}}\boldsymbol{A}_{ij}^{(m,m'')}\cdot\Big(\boldsymbol{\boldsymbol{\varepsilon}}_{jk}^{(m'',m')}\Big)^{[n-1]}\bigg].\label{eq:elastanceSolution}
\end{alignat}
Here $\boldsymbol{A}_{ik}^{(m,m')}=\tfrac{1}{2}\boldsymbol{I}_{ik}^{(m,m')}-\boldsymbol{L}_{ik}^{(m,m')}$,
and the primed summation in Eq.(\ref{eq:elastanceSolution}) indicates
that the diagonal term ($i=j=k$ and $m=m'=m''$) is excluded \cite{saad2003iterative}.
To start the iteration, we use the one-body solution of the linear
system, 
\[
\Big(\boldsymbol{\varepsilon}{}_{ik}^{(m,m')}\Big)^{[0]}=\delta_{ik}\delta_{mm'}\varepsilon_{ii}^{(l)}=\frac{\delta_{ik}\delta_{mm'}}{4\pi bDw_{m}}\mathbf{\Delta}^{(m)}.
\]
Chemical interactions, given by the off-diagonal ($i\neq j$) elastance
tensors, appear at the second iteration,
\begin{alignat*}{1}
\Big(\boldsymbol{\varepsilon}_{ik}^{(m,m')}\Big)^{[1]} & =-b^{m+m'}\bm{\nabla}_{{\scriptscriptstyle \boldsymbol{R}_{i}}}^{m}\bm{\nabla}_{{\scriptscriptstyle \boldsymbol{R}_{k}}}^{m'}H(\boldsymbol{R}_{i},\boldsymbol{R}_{k}).
\end{alignat*}
The solution of the elastance tensors at the third iteration is
\begin{alignat*}{1}
\Big(\boldsymbol{\varepsilon}_{ik}^{(m,m')}\Big)^{[2]} & =\Big(\boldsymbol{\varepsilon}_{ik}^{(m,m')}\Big)^{[1]}+\boldsymbol{\sum^{\prime}}\Big[-\tfrac{4\pi Db^{2m+2m'-1}}{(m'-1)!(2m'+1)!!}\bm{\nabla}_{{\scriptscriptstyle \boldsymbol{R}_{i}}}^{m}\bm{\nabla}_{{\scriptscriptstyle \boldsymbol{R}_{j}}}^{m'}H(\boldsymbol{R}_{i},\boldsymbol{R}_{j})\Big]\bm{\nabla}_{{\scriptscriptstyle \boldsymbol{R}_{j}}}^{m}\bm{\nabla}_{{\scriptscriptstyle \boldsymbol{R}_{k}}}^{m'}H(\boldsymbol{R}_{j},\boldsymbol{R}_{k}).
\end{alignat*}
The above recipe can then used to be used to systematically obtain
the elastance tensors at all orders. The higher order Jacobi solutions
are obtained in terms of higher gradients of the Green's function,
and are thus, subleading to the lower order solutions. In this work,
we have used the solution of the elastance tensor at the second iteration.
The coupling tensors of Eq.(\ref{eq:coupling}) satisfy
\begin{gather*}
\mathbb{\boldsymbol{\chi}}^{(l,m)}=-\frac{\tilde{w}_{l-1}w_{m}}{bw_{l+m-2}}\mathbf{M}_{i}^{(l+m-2)}\boldsymbol{I}+\frac{\tilde{w}_{l-1}w_{m}}{bw_{l+m}}\mathbf{M}_{i}^{(l+m)},
\end{gather*}
where the first term is non-vanishing for $l+m-2\geq0$, and $\mathbf{M}_{i}^{(m)}=w_{l}\int\mu_{c}(\boldsymbol{R}_{i}+\boldsymbol{\rho}_{i})\mathbf{Y}^{(m)}(\boldsymbol{\rho}_{i})\,dS_{i}$
are tensorial spherical harmonic coefficients of phoretic mobility.

\section{Simulation Details\label{app:Simulation-Details}}

With vanishing inertia, Newton's equations Eq.(\ref{eq:forceBalance})
can be inverted to obtain the rigid body motion of active colloids
in terms of known quantities:\begin{subequations}
\begin{alignat}{1}
\mathbf{V}_{i} & =\boldsymbol{\mu}_{ik}^{TT}\cdot\mathbf{F}_{k}^{\mathcal{P}}+\boldsymbol{\mu}_{ik}^{TR}\cdot\mathbf{T}_{k}^{\mathcal{P}}+\boldsymbol{\pi}_{ij}^{(T,l)}\cdot\negmedspace\boldsymbol{\chi}^{(l,m')}\negmedspace\cdot\boldsymbol{\varepsilon}{}_{jk}^{(m',m)}\cdot\mathbf{J}{}_{k}^{(m)},\\
\boldsymbol{\Omega}_{i} & =\boldsymbol{\mu}_{ik}^{RT}\cdot\mathbf{F}_{k}^{\mathcal{P}}+\boldsymbol{\mu}_{ik}^{RR}\cdot\mathbf{T}_{k}^{\mathcal{P}}+\boldsymbol{\pi}_{ij}^{(T,l)}\cdot\negmedspace\boldsymbol{\chi}^{(l,m')}\negmedspace\cdot\boldsymbol{\varepsilon}{}_{jk}^{(m',m)}\cdot\mathbf{J}{}_{k}^{(m)}.
\end{alignat}
\end{subequations}Here the propulsion tensors $\boldsymbol{\pi}^{(\alpha,l'\sigma')}$
give active contributions due to the slip \cite{singh2015many} and
the mobility matrices $\boldsymbol{\mu}^{\alpha\beta}$, with $(\alpha,\beta)=(T,R)$
give passive hydrodynamic interactions \cite{mazur1982,ladd1988}.
The above dynamical system, in simulations, is truncated at $l=3$
and integrated numerically using the open-source PyStokes library
\cite{pystokes} with an initial condition of random packing of hard-spheres
\cite{skoge2006packing}. We then study the system near a plane wall
by computing the mobility, propulsion and elastance tensors using
the Green's function of sec \ref{sec:minimalModel}. Our system is
only confined by a plane wall and there is no periodic boundary condition.
Thus, it is a distinctive feature of the present approach to colloid
hydrodynamics that a finite number of colloids in an infinite expanse
of fluid can be simulated, without having to either truncate the size
of the system or impose periodic boundary condition. 

\textcolor{black}{The orientations of the colloids are stabilized
along the wall normal by external torques $\mathbf{T}^{\mathcal{P}}=T_{0}(\mathbf{\hat{z}}\times\boldsymbol{p}_{i})$}.
The number of colloids $N$, for respective plots, are: Fig.3(a-d):
$N=1$; Fig.3(e-f): $N=2$; Fig.4: $N=2^{11}$. Other parameters used
in the simulations are: radius of colloids ($b=1$), strength of the
chemical monopole ($J_{0}^{(0)}=4$), strength of the chemical dipole
($J^{(1)}=10$), the strength of the bottom-heaviness ($T_{0}=0.2$),
diffusion constant ($D=1000$), and dynamic viscosity ($\eta=1$).
We vary the ratio $\alpha=J^{(1)}/J_{0}^{(0)}$ in Fig.4(n) to map
the state diagram. The conservative inter-particle force $\mathbf{F}_{i}^{\mathcal{P}}=-\boldsymbol{\nabla}_{{\scriptscriptstyle \mathbf{R}_{i}}}U$
is due to a short-ranged repulsive potential $U(r)=\epsilon\left(\frac{r_{min}}{r}\right)^{12}-2\epsilon\left(\frac{r_{min}}{r}\right)^{6}+\epsilon,$
for $r<r_{min}$ and zero otherwise \cite{weeks1971role}, where $\epsilon$
is the potential strength. The WCA parameters for particle-particle
repulsion are: $r_{min}=4.4,\,\epsilon=0.04$, while for the particle-wall
repulsion we choose $r_{min}=3,\,\epsilon=0.08$.\textcolor{black}{{}
}\end{appendix}

\end{document}